\documentclass[useAMS,usenatbib,fleqn]{mnras}			
\usepackage{times}
\usepackage{natbib} 
\usepackage[]{graphicx,color}
\usepackage{amsmath}
\usepackage{amssymb}
\usepackage{textcomp}
\usepackage[normalem]{ulem}
\usepackage{rotating}

\pdfoutput=1

\title[Spectroscopic decomposition of NGC~3311]
{Spectroscopic decomposition of the galaxy and halo of the cD galaxy NGC~3311}
\author[E. J. Johnston et al]{Evelyn~J.~Johnston$^{1}$\thanks{Email: ejohnston@astro.puc.cl}, Michael Merrifield$^2$, \& Alfonso Arag\'on-Salamanca$^2$\\
  $^1$European Southern Observatory, Alonso de C\'ordova 3107, Casilla 19001, Santiago, Chile\\ 
  $^2$School of Physics and Astronomy, University of Nottingham, University Park, Nottingham, NG7 2RD, UK\\ 
}

\begin{document}

\maketitle

\begin{abstract}
Information on the star-formation histories of cD galaxies and their extended stellar haloes lie in their spectra. Therefore, to determine whether these structures evolved together or through a two-phase formation, we need to spectroscopically separate the light from each component. We present a pilot study to use \textsc{buddi} to fit and extract the spectra of the cD galaxy NGC~3311 and its halo in an Integral Field Spectroscopy datacube, and carry out a simple stellar populations analysis to study their star-formation histories. Using MUSE data, we were able to isolate the light of the galaxy and its halo throughout the datacube, giving spectra representing purely the light from each of these structures. The stellar populations analysis of the two components indicates that, in this case, the bulk of the stars in both the halo and the central galaxy are very old, but the halo is more metal poor and less $\alpha$-enriched than the galaxy. This result is consistent with the halo forming through the accretion of much smaller satellite galaxies with more extended star formation. It is noteworthy that the apparent gradients in age and metallicity indicators across the galaxy are entirely consistent with the radially-varying contributions of galaxy and halo components, which individually display no gradients. The success of this study is promising for its application to a larger sample of cD galaxies that are currently being observed by IFU surveys.

\end{abstract}


\section{Introduction}\label{sec:introduction}

Galaxies display a wealth of morphologies which have been created over time through variations in the star-formation histories, and thus formation mechanisms, of their individual components \citep{Seigar_1998, MacArthur_2003, Allen_2006, Cameron_2009, Simard_2011}. cD galaxies are giant elliptical galaxies surrounded by an extended stellar halo, often found in the centres of clusters \citep{Mathews_1964, Schombert_1988, Whiley_2008,Zhao_2015b}. Their light profiles generally follow a $r^{1/4}$ profile in the inner regions, with an excess of light in the outskirts due to the halo \citep{Schombert_1986, Jordan_2004, Liu_2005}. The origin of these stellar haloes remains unclear, though many current theories focus on the halo being built up around the already-formed galaxy, for example from the stars ejected during the rapid merging of galaxies in the early stages of the initial cluster collapse \citep{Merritt_1984}, or through accretion of stripped stars from galaxies falling into the cluster \citep{Gallagher_1972,Richstone_1976}, or even the accretion of stellar haloes of galaxy groups that fall into the cluster \citep{Rudick_2006}. Detailed studies of the halo light can help us distinguish between these ejection and accretion scenarios, but care must be taken to extract the light only from the halo and to avoid contamination from the light from the central galaxy component.

Many studies have used light profile fitting of photometric data to measure the spatial properties of each component to determine whether they formed separately or together. \citet{Seigar_2007} found their best fits to a small sample of cD galaxies in the R-band when they used a S\'ersic-plus-exponential profile instead of a single S\'ersic profile, which they suggested was due to the outer stellar haloes forming independent of the galaxy through accretion, thus creating a second distinct component.

In another recent study, \citet{Zhao_2015a} fitted the $R$-band light distribution of a sample of 625 low-redshift brightest cluster galaxies (BCGs), and found that the differences in the effective radii ($R_{\rm e}$) for each component in a double-S\'ersic fit could be used to distinguish between elliptical and cD galaxies. Comparisons of the stellar masses and environment densities for the BCGs and cDs led them to the conclusion that cD galaxies originated as normal ellipticals which have grown in mass and size due to mergers, and later developed the stellar halos \citep{Zhao_2015b}. Furthermore, \cite{Zhao_2017} claim that from $z\sim2$ to $z\sim1$ star formation and merging contribute approximately equally to BCG mass growth, while merging plays a dominant role in BCG assembly at $z<1$. These authors also claim that the high-$z$ progenitors of present-day BCGs are not significantly different from other galaxies of similar mass at the same epoch, suggesting that the processes which differentiate BCGs from normal massive elliptical galaxies must occur at $z < 2$.

However, \citet{Bender_2015} showed that one must be careful when using photometric data alone to distinguish between the central elliptical and its extended stellar halo. Their study of NGC~6166 in Abell~2199 found that the cD galaxy and halo were best fit by a single S\'ersic profile with S\'ersic index 8.3, while the velocity dispersion increased with radius until it matched the mean cluster velocity dispersion, indicating that the halo has been built up through the accretion of stars from satellite galaxies through tidal encounters.

Detailed information on the star-formation histories of cD galaxies and their haloes lie in their spectra. Changes in age, metallicity and $\alpha$-enhancement between the central galaxy and its halo would indicate that the system was created through a two-stage formation mechanism where the inner regions formed in-situ while the outer regions were built up through a series of minor mergers \citep{Abadi_2006, Oser_2010}. \citet{Bender_2015} proposed that a super-solar [$\alpha$/Fe] ratio for the central elliptical (indicative of a relatively short star-formation period) and a near-solar [$\alpha$/Fe] for the halo (due to more extended star formation) would be an indicator that the halo was built up through accretion of tidal debris. A super-solar [$\alpha$/Fe]  ratio in the halo would also support this scenario, but would trigger new questions with regards to understanding how star formation in all the galaxies that contributed to the final galaxy must have shut off very early in their lifetimes.

Gradients in the age, metallicity and $\alpha$-enhancement observed across a galaxy can be used to constrain the stellar populations present in each component, and thus determine how they formed. Studies of elliptical galaxies with long-slit spectra have revealed that the metallicity generally drops as you move from the centre of the galaxy into the halo, while the stellar populations are generally old throughout and show super-solar $\alpha$-enhancement at the centre with a flat or negative gradient finishing with solar values in the halo \citep{Gorgas_1990,Cardiel_1995,Weijmans_2009, Pu_2010, Coccato_2010, LaBarbara_2012,Greene_2012}. Such gradients across the system can be explained either by a smooth transition as you move from the galaxy out into the halo, or by the superposition of two non-radially varying populations. The former scenario may suggest that the galaxy and halo formed together from the same material but over different timescales, while the latter scenario could reflect the two-phase formation scenario. 

In order to distinguish between these two scenarios, the spectral information of each component must be extracted cleanly. We have developed a technique called Bulge--Disc Decomposition of IFU data \citep[\textsc{buddi}, ][]{Johnston_2017}, which uses the photometric and spectroscopic information from the entire datacube to extract the one-dimensional integrated spectra of different large-scale structures within a galaxy. \textsc{buddi} was originally developed to model bulges and discs of galaxies. However, in fact it is more general and can be used to derive spectra for any photometrically-distinct components in a galaxy that can be modelled with \textsc{galfit}.

To explore the capabilities of \textsc{buddi}, we will examine the interrelationship between elliptical galaxies and their surrounding stellar halos in cD systems. NGC~3311, the cD galaxy at the centre of the Hydra I Cluster, provides an excellent target for this pilot study since is has been the focus of many previous studies using more traditional analyses with long-slit spectra, such as \citet{Ventimiglia_2010}, \citet{Coccato_2011b}, \citet{Arnaboldi_2012} and \citet{Barbosa_2016}, which can be used to compare with the results from this new analysis technique.

\section{Observations and Data Reduction}\label{sec:DR}
The Multi-Unit Spectroscopic Explorer \citep[MUSE, ][]{Bacon_2010} at the Very Large Telescope (VLT) is an optical wide-field integral-field spectrograph with a field of view of 1~arcmin$^2$, a spatial resolution of 0.2~arcsec/pixel, and a spectral resolving power ranging from \textit{R}$\sim$1770 at 480~nm to \textit{R}$\sim$3590 at 930~nm, making it perfect for the light-profile fits and spectral analysis required by this study.

\begin{figure*}
  \includegraphics[width=1\linewidth]{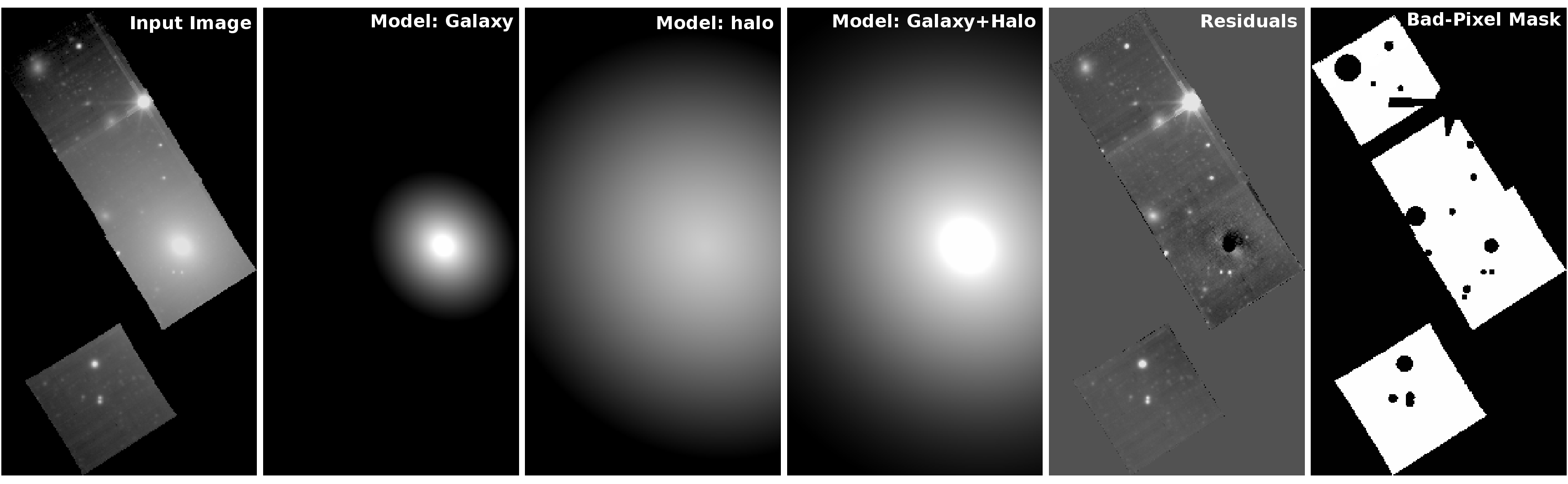}
  \caption{From left to right are the original white-light image for the datacube, the best fit model to the central galaxy, the best fit model for the halo, the combined galaxy+halo models, the residuals after subtracting the best fit model from the original image (with the area outside of the MUSE pointings masked out), and the bad pixel mask. North is upwards and East is to the left, and each image covers a $143\times 262\,$arcsecond$^2$ field-of-view.
 \label{fig:Galfit}}
\end{figure*}

MUSE observations of NGC~3311 were found to be publicly available on the ESO Science Archive Facility. The data were obtained between 2014--11--27 and 2015--01--21 as part of ESO Program 094.B-0711(A) (PI: Arnaboldi). The observations consist of a set of three pointings stretching north-east from the centre of NGC~3311 towards the halo, and a fourth pointing south-east of the centre of the galaxy, as shown in the left panel of Fig.~\ref{fig:Galfit}. Exposure times ranged from 670 to 1370 seconds, with four exposures per pointing, and each pair of exposures has a position angle rotation of $90\,$degrees to reduce the appearance of the slicers and channels in the final combined datacube. Due to the large size of the galaxy and the small amount of uncontaminated sky background in each pointing, a separate sky exposure was observed alongside each set of science observations for sky subtraction. Additionally, a standard star was observed on each night for flux calibration, sky flats were taken during twilight within 7 days, and internal lamp flats were taken with the calibration lamps at the start of each set of observations to account for the time-dependent, temperature-related variations in the flux levels between each IFU. While such deep observations of this galaxy are not strictly necessary for the analysis that we wish to undertake here, it will allow us to perform a more traditional study of the datacube for comparison which will provide further confidence in applying this technique to lower-quality observations.

The data were reduced using the ESO MUSE pipeline \citep{Weilbacher_2012} in the ESO Recipe Execution Tool (EsoRex) environment \citep{ESOREX}. Master bias and flat-field frames and wavelength solutions were created for each night of observations using the associated raw calibrations, and used to apply a basic data reduction to the science and standard star observations. The standard star and dedicated sky frames were then processed to obtain the flux calibration solution and the sky continuum, which were applied to the science frames as part of the post-processing step. Finally, the reduced pixel tables created from the post-processing step for each exposure were combined to produce the final datacube. The data was reduced in this way within the wavelength range of 4780\AA\ to 7220\AA\ since this region covered the spectral features of interest for this work.

\section{Fitting the Galaxy and Halo}\label{sec:method}

An IFU datacube can effectively be seen as a series of images of the target at each wavelength step. If the field-of-view is large enough with sufficient spatial resolution and the signal-to-noise ratio (S/N) of the data set is high enough,  conventional fitting and decomposition techniques can be applied to each image to determine the fraction of light and the properties of each component at that wavelength. Application of this technique to every image slice in the data cube can thus provide the luminosity fraction from each component as a function of wavelength, otherwise known as its spectrum.

However, this approach is not feasible where the S/N is too low or where the galaxy contains multiple components of different colours, and is especially problematic at wavelengths where there is strong contamination from sky lines. In such cases, one needs to use all the information contained within the datacube simultaneously to find the best wavelength-dependent model for the galaxy. For this study, the galaxy fitting was carried out using \textsc{buddi}. \textsc{buddi} uses a modified form of \textsc{Galfit} \citep{Peng_2002, Peng_2010} called \textsc{GalfitM} \citep{Haeussler_2013, Vika_2013}, which is able to read in and fit multi-waveband images simultaneously. \textsc{GalfitM} uses user-defined chebychev polynomials to fit any variations in the structural parameters of each component smoothly over the wavelength range, thus using information from all the images for each fit, boosting the S/N of the data set over that of any individual image and allowing reasonable estimates to be made for images with lower S/N.

A full description of \textsc{buddi} can be found in \citet{Johnston_2017}. In the following sections we give a brief overview.

\subsection{Preparing for the fits}\label{sec:considerations}
Before starting the fitting process with \textsc{buddi}, a few preparations must be made. The first of these preparations was to create a series of reliable bad pixel masks for each white-light image, narrow-band image and image slice used in the fits to ensure that \textsc{galfitm} was only using information for the galaxy light. The first step towards creating these masks was to produce a bad pixel datacube with the same dimensions as the original datacube, with those spaxels outside of the MUSE pointings masked out. Distortions to the fits can also occur due to the presence of additional light sources nearby, such as foreground stars and background galaxies. The bad pixel datacube was correspondingly updated to mask out these additional sources, along with the diffraction spikes created by the bright star to the north of the galaxy in the MUSE pointings that can be seen in Fig.~\ref{fig:Galfit}. Additionally, the datacube reveals a central emission-line region that is also affected by dust \citep{Vasterberg_1991}, which distorts the best fit. Therefore, a circular region of radius 4~arcseconds was also included in the mask datacube. The datacube was then manipulated in the same way as all the other images created by \textsc{buddi}, resulting in a series of bad pixel masks corresponding to each image used in the fits. An example of the bad pixel mask for the white-light image can be seen on the right of Fig.~\ref{fig:Galfit}.

Another element to be considered is that the fitting process requires each model to be convolved with a point-spread function (PSF) profile. A PSF datacube was created using data from the unsaturated foreground stars in the original datacube, which were cleaned, recentred and combined by median superposition. The PSF datacube was then binned in the same way as the science datacube to ensure that each white-light image, narrow-band image and image slice had a corresponding PSF profile at the same wavelength.

A final factor to take into account is that the sky subtraction carried out by the ESO MUSE pipeline is not yet optimal, leading to the presence of artefacts in the final datacube due to the parametrization of
the line spread function. Therefore, to take into account small variations in the background sky levels due to this issue, a residual-sky component was included in the models that was constant across each image but allowed full freedom to vary in the wavelength direction. Simulations of the effect of the sky background in single-S\'ersic fits by \citet{Haeussler_2007} have shown that as long as the galaxy light profile is well approximated by the model used, the sky background values given by \textsc{galfit} can be considered a good approximation. \citet{Johnston_2017} have further shown that \textsc{galfitm} was successful at reducing the sky contamination in the decomposition of MaNGA datacubes, though it doesn't remove the effect completely.

\subsection{Step 1: Obliterate the kinematics}\label{sec:step1}
\begin{figure}
\centering
  \includegraphics[width=.99\linewidth]{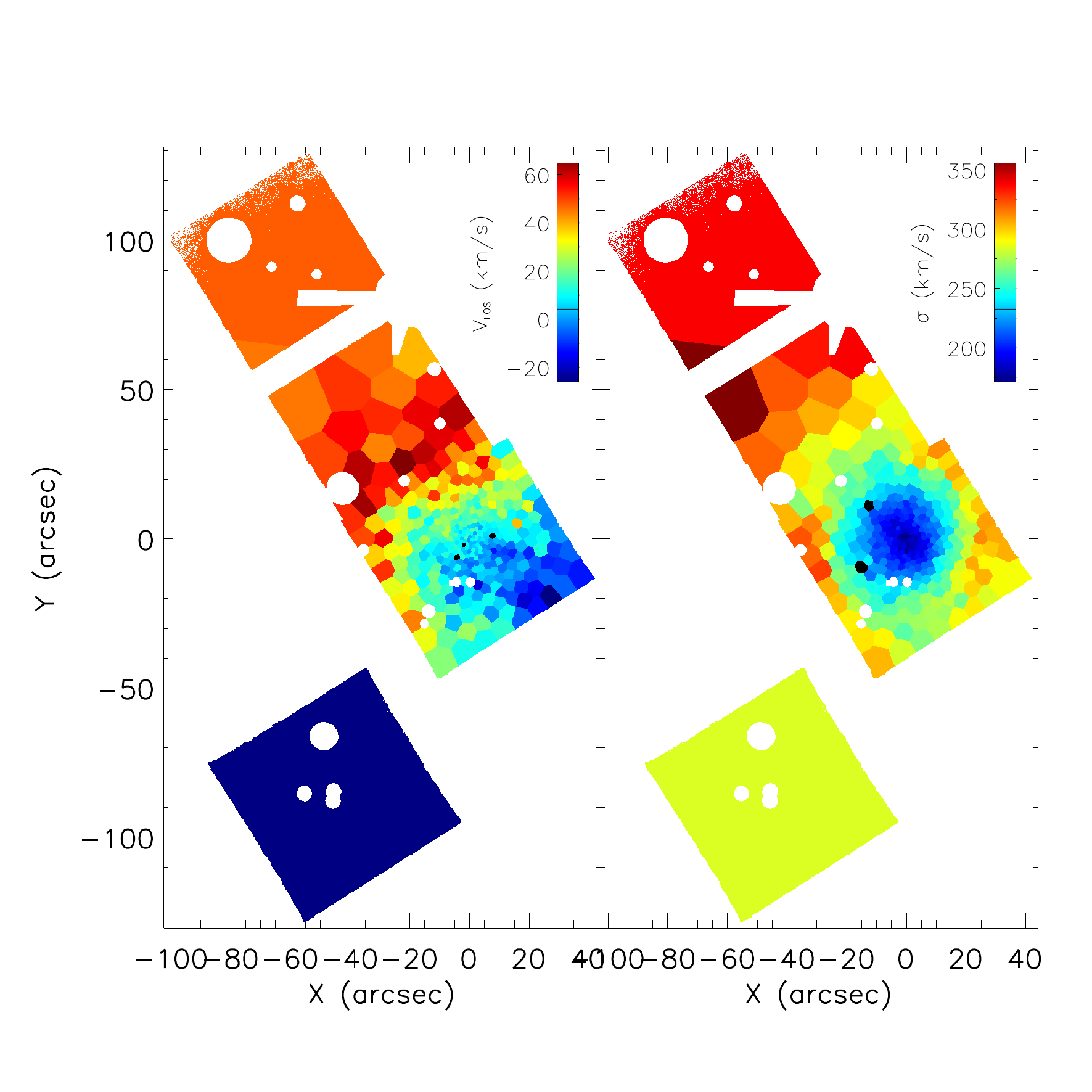}
\caption{Maps of the line-of-sight velocity normalised to the velocity at the centre of the galaxy (left) and velocity dispersion (right) measured across NGC~3311, which were obliterated as the first step of the fitting process. The white points show the regions masked out in the fit, either due to the presence of a foreground or background object that would affect the measurement or lying outside of the MUSE data field of view.}
\label{fig:kinematics}
\end{figure}

The first step towards modelling the galaxy with \textsc{buddi} is to measure and normalise the kinematics across the galaxy. The datacube was first binned using the Voronoi tessellation technique of \citet{Cappellari_2003}, and the kinematics measured in the wavelength region \mbox{4800~$\le \lambda \le$~5900\AA} with the penalised Pixel Fitting software (\textsc{pPXF}) of \citet{Cappellari_2004}. pPXF combines a series of stellar template spectra of known ages and metallicities, and convolves them with a range of line-of-sight velocity distributions (V$_{\text{LOS}}$), velocity dispersions ($\sigma$) and  Gauss-Hermite polynomials (h$_3$ and h$_4$), to produce a model spectrum that best fits the binned galaxy spectrum and its kinematics. For this study, the MILES stellar library \citep{Sanchez_2006} was used, consisting of 156~template spectra ranging in metallicity from -1.71 to +0.22, and in age from 1 to 17.78~Gyrs. This step is necessary since \textsc{Galfit} can only fit symmetric models, which would not be adequate for IFU image slices of a rotating galaxy, especially close to a strong spectral feature where one part of the galaxy represents the continuum light while another part is fainter since it is in absorption due to the rotation-induced Doppler shift of the spectral lines. Therefore, the line-of-sight velocity and velocity dispersion across the datacube are normalised such that each image slice shows the galaxy at the same rest-frame wavelength and with the same spectral broadening across the full field-of-view to avoid asymmetries. Figure~\ref{fig:kinematics} displays the maps of the line-of-sight velocity and velocity dispersion measured across NGC~3311, where the white areas show the regions masked out during the binning. The mask is the same as the bad-pixel mask, but with the centre of the galaxy unmasked. The kinematics corrections for the masked spaxels within the MUSE pointings were set to the same value as the nearest binned spaxels. An additional benefit of this kinematics-obliteration step is that if different components within a galaxy have slight differences in their line-of-sight velocities, the smoothing applied by \textsc{buddi} will smear out these differences, reducing the effect on the final spectra.

As noted by earlier studies \citep[see e.g.][]{Hau_2004, Ventimiglia_2010, Richtler_2011}, the velocity dispersion was found to increase from $\sim180\,$km/s in the centre of the galaxy to $\sim360\,$km/s at a radius of $50\,$arcsec, thus approaching the velocity dispersion of the cluster of $784\,$km/s \citep{Misgeld_2008} but not reaching it within the field of view of the data. As can be seen in Fig.~\ref{fig:kinematics}, the same trend is seen in this data set. Therefore, the data were corrected for the variations in the line-of-sight velocity over the galaxy due to its rotation, and broadened to 355~km/s to match the maximum velocity dispersion in the stellar halo within the field of view of the MUSE data.

\subsection{Step 2: Fit the white-light image}\label{sec:step2}
Due to the huge amount of CPU time that would be required to fit all images in a MUSE datacube simultaneously, \textsc{buddi} carries out the fit using a multi-step process. The first step is to create and fit a white-light image of the datacube, which allows the user to quickly determine an approximate best fit model for the galaxy, including the number of components necessary for a good fit. 

NGC3311 was first modelled with a single S\'ersic profile, but many artefacts were found to remain in the residual images, indicating that an additional component was needed for a good fit. Acceptable fits were found using 2-component models, with little improvement seen when a third component was added. We therefore stuck with a 2-component model representing the central galaxy and extended outer halo. To assess the sensitivity of the fitting process to the functional form adopted, the datacube was fitted twice with two different 2-component models with \textsc{buddi}. Model~1 uses a typical de Vaucouleurs plus exponential profile for the elliptical galaxy and its extended stellar halo respectively \citep[]{deVaucouleurs_1953, Seigar_2007}, while Model~2 consists of a double S\'ersic profile to allow additional flexibility in the fit. An example of the fit with the double-S\'ersic model can be seen in Fig.~\ref{fig:Galfit}, showing the original white-light image from the datacube, the galaxy and halo models, the best fit model,  and the residual image after subtraction of the best-fit model. It can be seen that even though the MUSE data only cover a limited section of the galaxy, \textsc{GalfitM} has been able to converge on a good solution, with the residuals only displaying significant flux in the masked regions. This statement is true for the application of both models to the white-light image, and at this stage is was difficult to identify which model best represented the system. 

It should be noted at this point that NGC~3311 is not a simple galaxy+halo system as represented by the models used here. Previous works that have applied similar fits to photometric data of this system found evidence of satellite galaxies, tidal-tails and an off-centre halo \citep{Arnaboldi_2012, Barbosa_2018}, all of which can affect the fit. Therefore, the fits presented here represent the system only within the field-of-view of these observations, and assume that the light profile outside of this field follows the same trend.

\begin{figure*}
\centering
  \includegraphics[width=.9\linewidth]{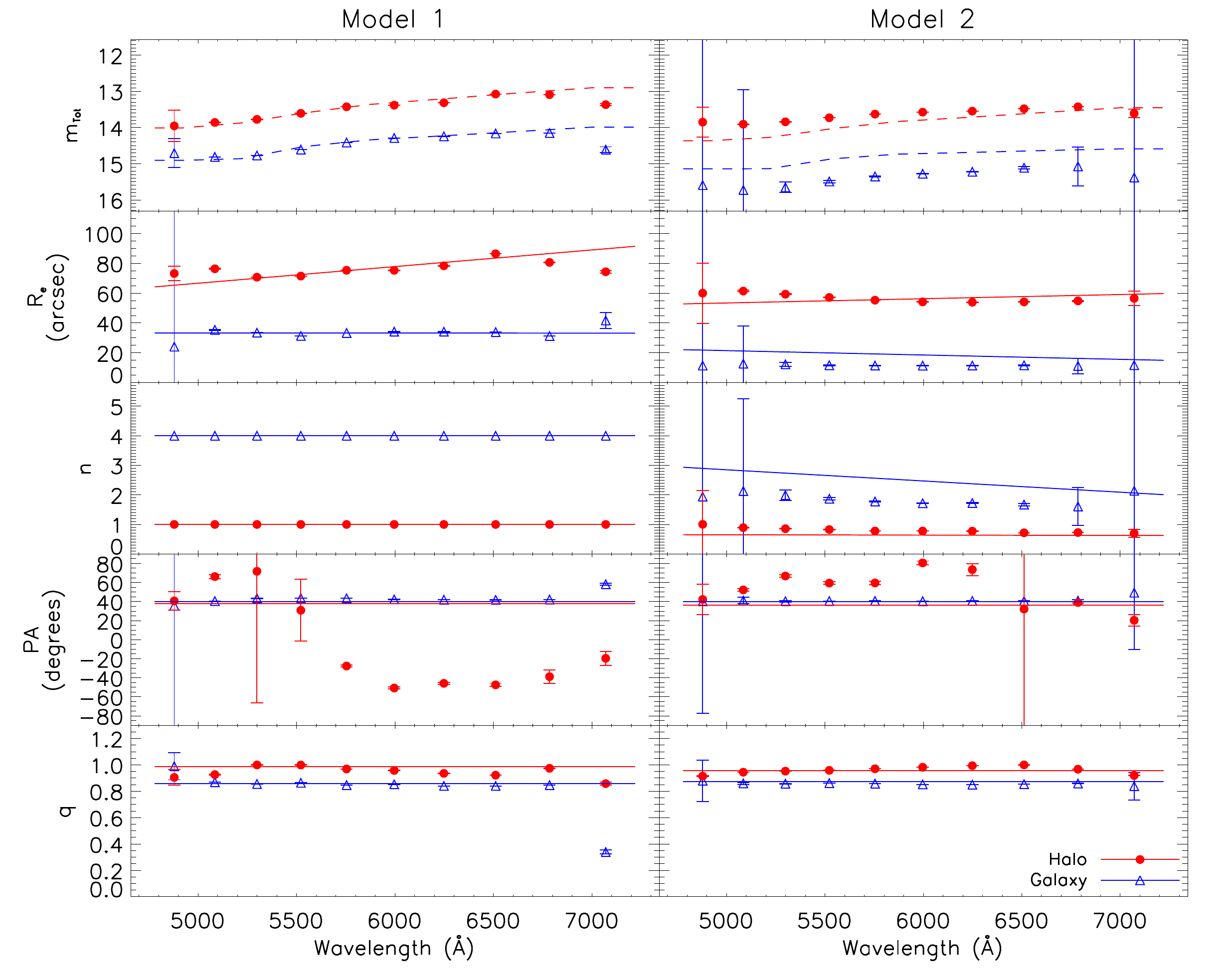}
\caption{Variations in physical parameters as a function of wavelength for the fits to NGC~3311 using Model~1 (left) and Model~2 (right). Dots represent values from the independent fits to narrow-band images, and lines represent the best fit when modelled simultaneously. Measurements in red and blue correspond to the halo and galaxy respectively. Solid lines show the constrained fits that are applied to the fits to individual image slices in the final fitting step, while dashed lines show the parameters that are allowed complete freedom to vary in those fits. }
\label{fig:parameters}
\end{figure*}

\subsection{Step 3: Fit the binned/narrow-band images}\label{sec:step3}
The next step is to refine the best fit parameters over the wavelength range of the datacube and to allow the introduction of smooth variations in the parameters as a function of wavelength to represent colour gradients within the galaxy or halo if present. This step is carried out by applying fits to a series of binned images, or `narrow-band'  images to use the terminology in \citet{Johnston_2017},  from across the datacube. The datacube was divided into 10 sections of equal size along the wavelength direction, and the images in each section were median binned to create a series of high S/N narrow-band images over the datacube. Since the datacube was logarithmically binned in wavelength, each narrow-band image was created from 190 image slices representing a wavelength range of $\sim$206\AA\ in the blue part of the spectrum to $\sim$285\AA\ at the red end. The use of the median values for each pixel across the set of images ensures that artefacts in the datacube, such as residual sky lines or dead pixels, are not apparent in the final narrow-band images and thus do not affect the fits to those images. Similarly, the widths of the spectral features in the datacube are typically significantly smaller than the wavelength range over which each narrow-band image was created, meaning that the presence of strong emission or absorption features within the wavelength range of the narrow-band image has a negligible effect on the final image.

\begin{figure*}
  \includegraphics[width=0.9\linewidth]{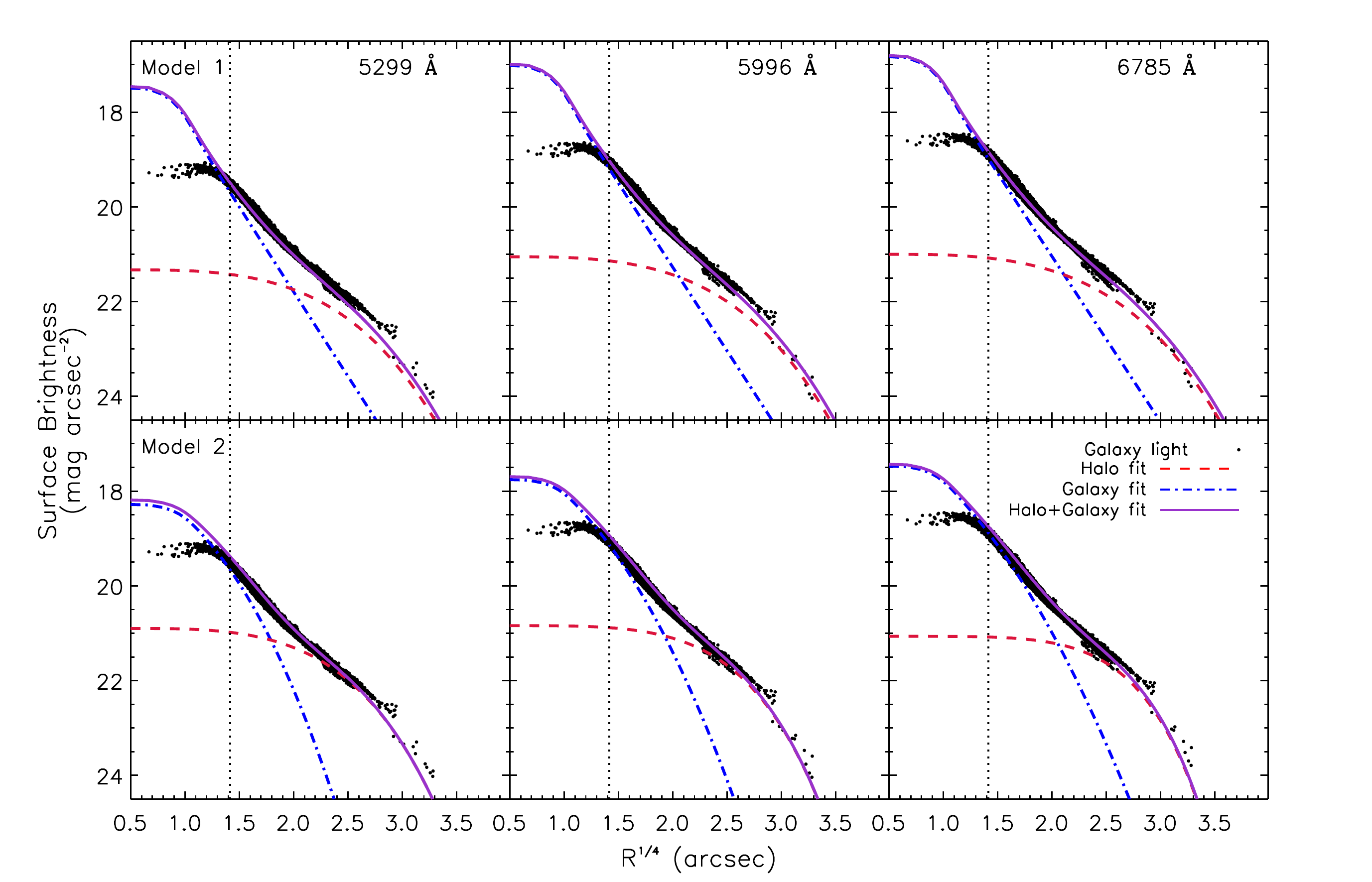}
  \caption{One-dimensional light profile fits for the de Vaucouleurs and exponential profile (Model~1, top) and the double S\'ersic profile (Model~2, bottom) for the narrow-band image centred at $5299\,$\AA , $5996\,$\AA\ and $6785\,$\AA (left to right).The vertical dotted line indicates the central circular region of radius $4\,$arcsec that was masked out of the fit (see text for details). 
 \label{Fig:light_prof}}
\end{figure*}

The binned images were first fitted independently with \textsc{galfitm} to obtain an estimate of how the physical parameters vary across the wavelength range. Due to the depth of the data used for this study and the increased S/N obtained by binning the data, the results from this step were found to be very smooth across the images, as shown by the red and blue points in Fig.~\ref{fig:parameters} representing the halo and galaxy respectively. The uncertainties on these measurements represent the statistical uncertainties calculated in the fit by \textsc{galfitm}. These uncertainties should be considered a lower limit since they are calculated using the assumption that the galaxy is well modelled with the number of components used in the fit and only contains Poissonian noise, a scenario which is very rare in reality \citep{Haeussler_2007}. 

The median value for each parameter from the independent fits was then taken as the starting value to fit the images simultaneously with \textsc{galfitm}, and the polynomial for the fit determined from the trends seen in the individual fits. In both models, the effective radii were constrained to follow a linear polynomial with wavelength, and the position angles and axis ratios were allowed to vary while remaining constant with wavelength. In Model~1, the S\'ersic indices were held fixed at $n~=~1$ for the halo and $n~=~4$ for the galaxy, while in Model~2 they were allowed to vary according to a linear polynomial. For both models, the integrated magnitudes (and thus total fluxes) of each component were allowed complete freedom. Figure~\ref{fig:parameters} shows that for both models, the constrained fits to the narrow-band images were found to be in good agreement with the results from the individual fits. At this step however one can start to see the effects of imposing additional constraints in the fit parameters in Model~1. For example, both the S\'ersic indices and effective radii of each component drops when allowed additional freedom in Model~2, leading to a reduction in the concentration of light at the centre of the model. The position angle of the halo in both models varies widely with wavelength. However, since \textsc{galfitm} perceives both components to be close to circular on sky ($q\sim1$), variations in the position angles of each halo are insignificant in Fig.~\ref{fig:parameters}. 

To further visualise the differences between these models, the one-dimensional light-profile fits for the narrow-band image slices centred on 5299\,\AA, 5996\,\AA\ and 6785\,\AA\ are shown in Fig.~\ref{Fig:light_prof}. One obvious difference between the two fits is the treatment of the light profile within the central masked region, where the more constrained de~Vaucouleurs plus exponential model shows a brighter core at all wavelengths. Additionally, in the black points representing the light-profile of the galaxy, one can see a divergence in the curve outside of 28~arcseconds (R$^{1/4}\sim$2.3). This trend reflects the asymmetric and offset halo, which is brighter to the north-east of the galaxy than the south-west, and the fit to the halo by \textsc{galfitm} clearly attempts to find a common solution between these two profiles.

\begin{figure*}
\centering
  \centering
  \includegraphics[width=.9\linewidth]{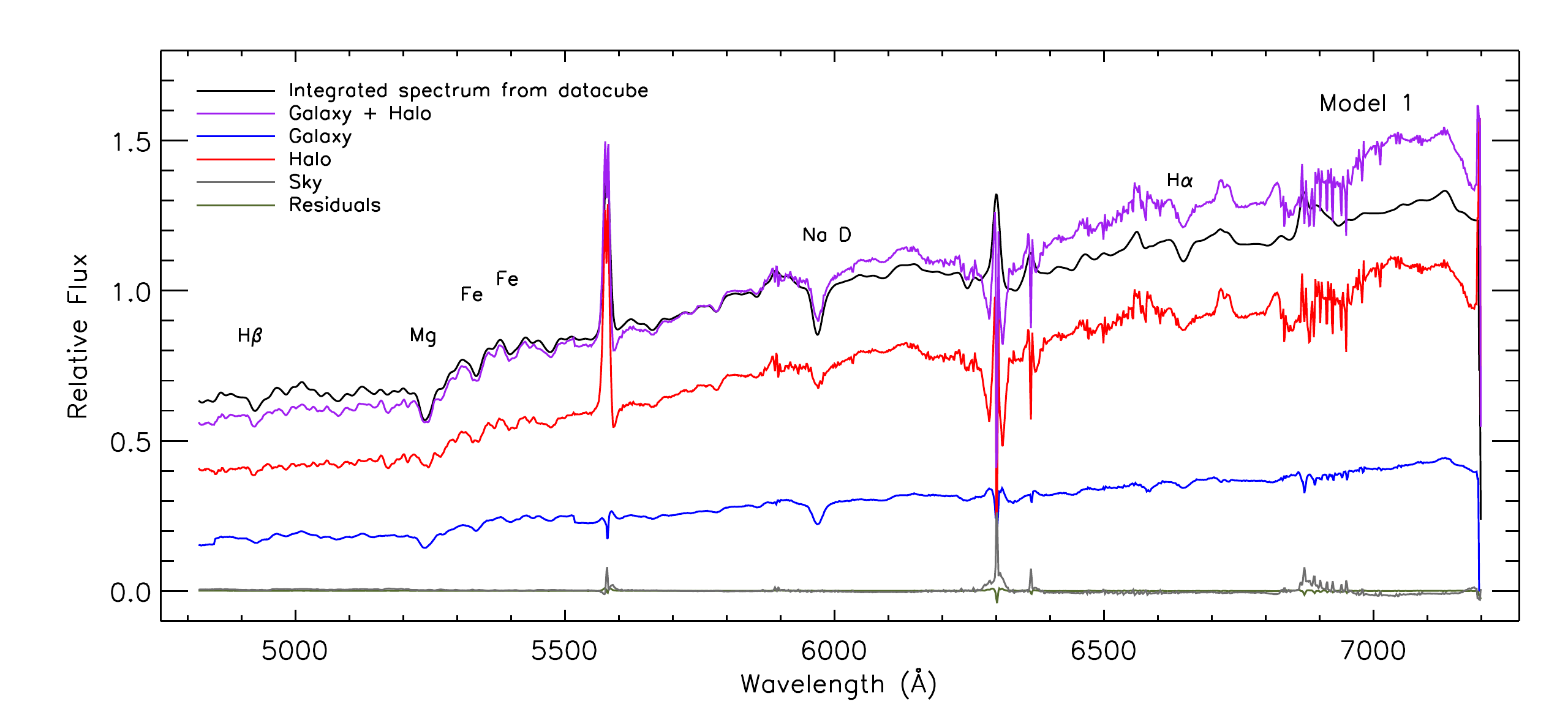}
  \centering
  \includegraphics[width=.9\linewidth]{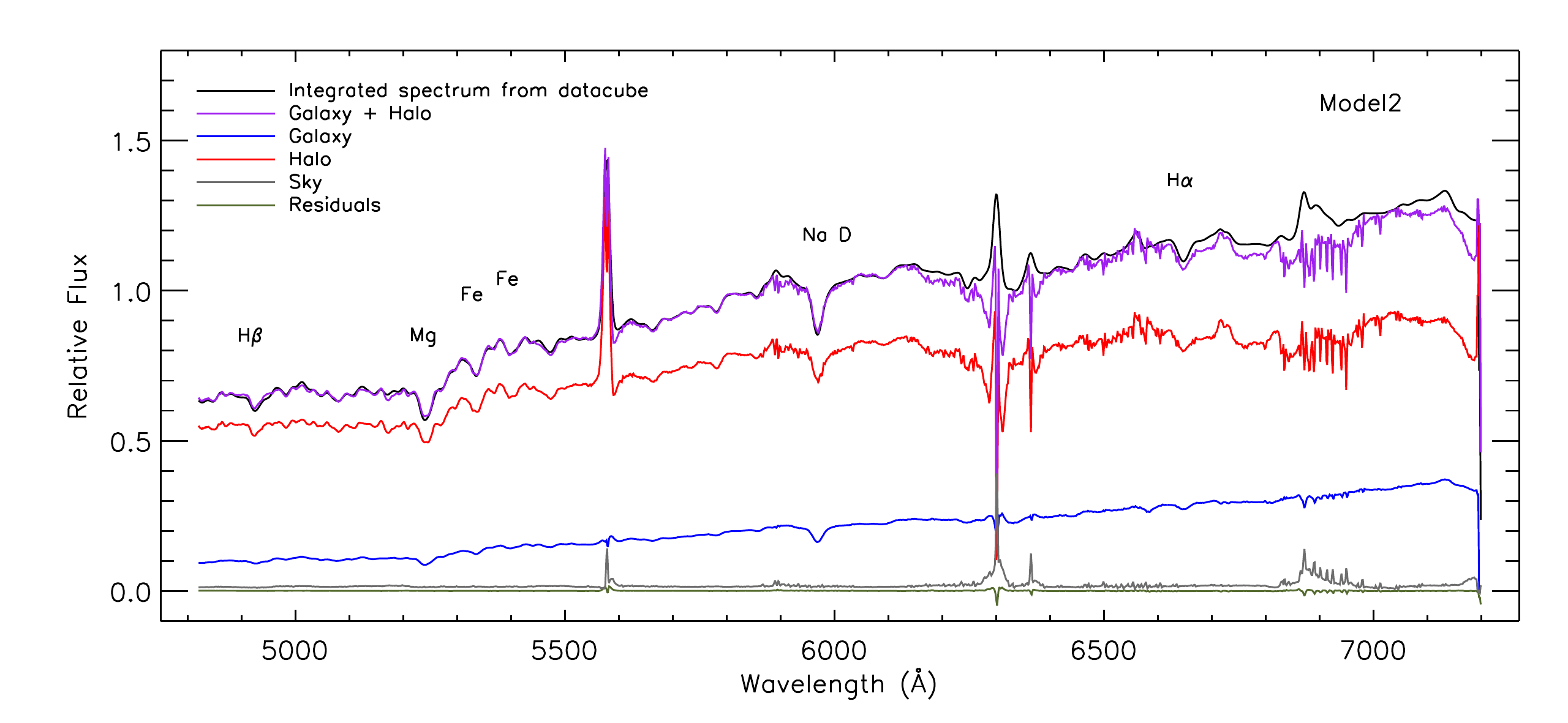}
\caption{A comparison of the decomposed spectra from Model~1 (top) and Model~2 (bottom) with the integrated spectrum from the original datacube in the rest-frame of the centre of the galaxy. The black spectrum is that of NGC~3311 integrated over the MUSE field of view after applying the kinematics-obliteration step, the red and blue spectra represent the model halo and galaxy spectra, and the purple spectrum is the the composite galaxy+halo spectrum. The residual sky spectrum measured by \textsc{galfitm} and the residual flux are shown by the grey and green lines. The model spectra were integrated to infinity, and the residual sky and residual spectra were integrated over the field-of-view of the datacube since \textsc{galfitm} gives these values as counts as opposed to integrated magnitudes. All these spectra were then normalised relative to the flux of the one-dimensional galaxy spectrum.}
\label{fig:spectra}
\end{figure*}

\subsection{Step 4: Fit the image slices and obtain the decomposed spectra}\label{sec:step4}
Finally, the datacube is split up into batches of ten image slices and a constrained fit applied to each image using the best fit parameters from the fit to the narrow-band images. The fits were constrained by holding the values for the effective radius, S\'ersic index, position angle and axis ratio fixed according to the best fit polynomial obtained in the previous step, while the integrated flux was allowed complete freedom to vary. In theory, any number of images can be used in each batch, however we find that keeping the number low, $\sim10-15$, results in a good balance between faster fits to each batch of images and reducing the CPU power needed to hold so many images and parameters in the system memory at any given time.

The one-dimensional decomposed spectra for each component can then be created from the total flux as a function of wavelength. Since \textsc{buddi} reduces a datacube to a few one-dimensional spectra, the resulting signal-to-noise ratios of the derived component spectra are very high, allowing them to be readily analysed in terms of their stellar populations. The decomposed spectra for both models are shown in Fig.~\ref{fig:spectra}, where the black line represents the integrated flux from the MUSE datacube after the kinematics-obliteration process outlined in Section~\ref{sec:step1}, and the blue, red and purple lines represent the light from the galaxy, the halo and the galaxy+halo respectively, all of which have been normalised to the median flux level of the integrated galaxy+halo flux. Interestingly, one can see in these plots how the SED of the model galaxy goes from being under- to over-estimated from the blue to red wavelength ranges due to the additional constraints imposed in Model~1, while the model galaxy spectral energy distribution (SED) from Model~2 follows closely that of the input data. Since these model spectra represent the total integrated light of each component when extrapolated to infinity and are being compared to the spectrum of the system from the original datacube integrated over the available field-of-view, it was considered likely that this offset was due to how the light profile was modelled by \textsc{galfitm} outside of the field of view of the data. To test this hypothesis, the light profile information of each model was used to create model datacubes for the galaxy and halo, and the one-dimensional spectra of each component computed by co-adding the flux in all pixels with valid data in the original datacube. These spectra were plotted in the same way for comparison, and it was found that in both models the fraction of light from the halo decreased, and the fit to the integrated spectrum from the original datacube was significantly improved in the case of Model~1. This improvement in the fit is shown by the residuals spectrum in Fig.~\ref{fig:spectra}, which was created by subtracting the best-fit galaxy+halo+sky datacube from the kinematic-obliterated datacube, and thus represents the residuals over the MUSE pointings only. 

After consideration of the shape of the SED, the main differences between the integrated model spectrum and the original datacube are due to the presence of strong sky lines, such as those at $\sim $5580\AA\ and $\sim$6300\AA\ in Fig.~\ref{fig:spectra}, and the lines between $\sim$6800--7000\AA\ which have been blended together during the kinematics obliteration step. The constraints in the fit parameters to the image slices are able to account for small variations in the sky background, as shown by the residual sky spectrum in Fig.~\ref{fig:spectra}, but lead to distortions when a strong sky background is present. In addition, it must be remembered that the sky lines do not have a rotational velocity like the galaxy, and so the shape of these spectral features integrated across the field-of-view is distorted during the kinematics-obliteration step of \textsc{buddi}. As a result, the sky background is no longer continuous over the image slices at these wavelengths, which further affects the fit.

At this stage, we can see that while both models looked comparable when fitting the narrow-band images, the double-S\'ersic model produced the decomposed spectra that best represent the original datacube when combined. This result is reminiscent of the warning of \citet{Bender_2015} that one must be careful when fitting the central elliptical galaxy and its extended stellar halo of a cD galaxy using the light profile information alone.

\section{Analysis of the fit parameters}\label{sec:analysis_fits}

The nature of the light profile fits to galaxies and their extended stellar haloes has been shown to reflect similarities and differences in their evolutionary histories. Several studies have found that the best fits to cD galaxies are obtained using two-component models as opposed to one-component models, and similarly that the differences in the two-component models reflect the different formation pathways. \citet{Gonzalez_2005} used a double de~Vaucouleur profile to fit the BCGs in 24 cluster galaxies, and found that while the properties of the inner component were clearly associated to the BGC, the outer component was consistent with the intracluster stars, suggesting that it had been built up through accretion of this material. Similarly, \citet{Seigar_2007} concluded that the double S\'ersic  and S\'ersic+exponential profiles that they used to fit their sample of cD galaxies represented the halo and central galaxies as distinct structures that formed through different processes. However, \citet{Bender_2015} found that in the case of NGC~6166, using photometry alone, the best fit was instead found with a single S\'ersic profile with a high S\'ersic index of 8.3, and that a kinematical decomposition was also needed to explain the increasing velocity dispersion with radius, and thus determine that the halo was built up around the galaxy through accretion of material from the cluster.

\begin{table}
\centering
\begin{tabular}{cccccc }
                        & \multicolumn{2}{c}{Model 1} &  & \multicolumn{2}{c}{Model 2} \\
                        & Galaxy       & Halo       &  & Galaxy       & Halo       \\ 
\hline

\multicolumn{1}{l|}{$R_{\rm e}$ }  & 33$\pm$4                   & 80$\pm$5          &  & 17.9$\pm$0.5            & 57$\pm$3          \\
\multicolumn{1}{l|}{$n$}                           & 4.00                            & 1.00                   &  & 2.38$\pm$0.19            & 0.64$\pm$0.09          \\
\multicolumn{1}{l||}{$q$}                       & 0.86$\pm$0.17        & 0.99$\pm$0.04  & & 0.872$\pm$0.011            & 0.96$\pm$0.03          \\
\multicolumn{1}{l|}{PA}                           & 40$\pm$6              & 38$\pm$50      &  &  40$\pm$3           & 36$\pm$20              \\ \hline 

\end{tabular}
\caption{The best-fit parameters for the effective radius in arcseconds ($R_{\rm e}$), S\'ersic index ($n$),  axis ratio ($q$), and position angle (PA) for the galaxy and halo measured in degrees with 0 pointing North, using a de Vaucouleurs and exponential profile (Model~1) and a double S\'ersic profile (Model~2). The uncertainties represent the standard deviation in the fit parameters for the binned images when modelled independently with \textsc{Galfit}.}
\label{Table:fit_results}
\end{table}

The best fit parameters for NGC~3311 at 6231\AA, the centre of the $r$-band, are given in Table~\ref{Table:fit_results} as calculated by interpolation from the best fit parameters to the narrow-band images. The errors represent the standard deviation in the fit parameters over all the binned images when fitted independently with \textsc{Galfit} (see Fig.~\ref{fig:parameters}). A direct comparison of these measurements with the literature is tricky due to the small field-of-view of the MUSE data compared to the size of the galaxy, meaning that the off-centre halo is not well displayed in the field and that the contamination from the companion galaxy, NGC~3309, is difficult to quantify. Previous works have modelled NGC~3311 as a single S\'ersic profile, such as \citet{Arnaboldi_2012} and \citet{Vasterberg_1991} who found effective radii of 270~arcsec ($n=5$, V-band) and 80~arcsec ($n=4$, $r$-band) respectively. However, recent fits to V-band photometry of NGC~3311 by \citet{Barbosa_2018} over a larger field of view found a best fit using a four-component model-- three components to fit the galaxy and a fourth to model the halo, in addition to modelling NGC~3309 and a neighbouring dwarf galaxy to reduce their effects on the fits. Multi-component fits to cD galaxies have also been found to best characterise their light profiles in other recent works \citep[e.g.][]{Huang_2013a, Spavone_2017}. However, due to the limitations outlined above with this MUSE data set, we stopped at a 2-component fit to reflect the galaxy+halo components since little noticeable improvement was found by incorporating additional components into the model.

As can be seen in Fig.~\ref{fig:parameters}, no significant gradient in the effective radius was detected with wavelength, suggesting that neither component contains a strong colour gradient \citep{Johnston_2012}. A possible gradient may exist in the effective radius of the halo for Model~1, but it is more likely that this gradient has been induced by the S\'ersic index constraint in that fit. Since colour gradients are often used as indicators of age and metallicity gradients, the flat gradients in the fit parameters therefore suggests that the galaxy and halo components contain flat or very shallow age and metallicity gradients. This finding is in agreement with the flat line-index gradients detected within 1~R$_e$ of NGC~3311 by \citet{Loubser_2012}.

The ratios of the effective radius of the galaxy to that of  the halo are 0.43$\pm$0.07 and 0.34$\pm$0.04 for Models~1 and 2 respectively, which are consistent with \citet{Seigar_2007} and \citet{Gonzalez_2005}, who found a range of $0.1$--$0.4$ for their samples of cD galaxies. Similarly, using the total flux of each component when extrapolated to infinity we find that the halo accounts for $\sim71$\% of the total light from the system in both models. This result is slightly outside of the estimates of \citet{Zhao_2015a} of $40$--$60$\%, but in agreement with \citet{Seigar_2007}, who measured $60$--$80$\% when they extrapolated their models to infinity, and lies toward the lower end of the range of luminosity fractions measured by \citet{Gonzalez_2005}. 

It must be noted at this point that the models we use for NGC~3311 in this study assume that the halo is smooth and symmetric both inside and outside of the available field of view. In reality however, \citet{Arnaboldi_2012} and \citet{Barbosa_2018} found that the halo is asymmetric, with an excess of light towards the north east of the central galaxy due to the halo lying off-centre and the presence of what that they identified as a tidal stream from an in falling galaxy. This effect can be seen in Fig.~\ref{Fig:light_prof}, where the measurements from the narrow-band image appear to diverge beyond $\sim$28~arcseconds, or R$^{1/4}\sim$2.3~arcseconds$^{1/4}$. As a result,  fits to these data have been affected such that our measurements for the physical properties of the halo are intermediate between those one would measure if both halves were fitted independently. While this effect would have little impact on the decomposed spectra if the halo stellar populations are uniform throughout, the presence of substructures within the halo due to accreted material would affect the spectra. Fits to imaging data by \citet{Arnaboldi_2012} and \citet{Barbosa_2018} have revealed three distinct substructures within the field-of-view of the MUSE data, which are attributed to tidal tails. As a result, if the stellar populations of these substructures are significantly different to that of the halo, it is likely that the decomposed halo spectrum obtained from this data will be biased towards these properties. One approach that would overcome such issues is to incorporate the non-parametric component capability in \textsc{galfitm} into \textsc{buddi}, which would allow such features to be detected during the fitting process and automatically be masked out. While this approach is planned in the future development of \textsc{buddi}, it is beyond the scope of this work. Our approach instead focussed on masking out only the light from bright foreground and background objects within the data. While the faint substructures are still present within the data, their light is less luminous than that of the halo and galaxy, and so the effect of this contamination should be minimal given the large area over which the fits are applied, resulting in a spectrum representing the mean global properties of the halo.  

Nevertheless, the light profile fits to NGC~3311 shown in figure~\ref{Fig:light_prof} clearly indicate the presence of two distinct components, which we have attributed to the central galaxy and extended halo, and it seems quite plausible that these components may have had separate formation histories. As other studies have suggested, the stellar halo may have been built up through accretion of smaller galaxies after the formation of the central galaxy. We will explore this possibility in the next section.

\section{Stellar Populations Analysis}\label{sec:stellar_pops}
\label{sec:stellarpop}

\begin{figure}
\includegraphics[width=0.95\linewidth]{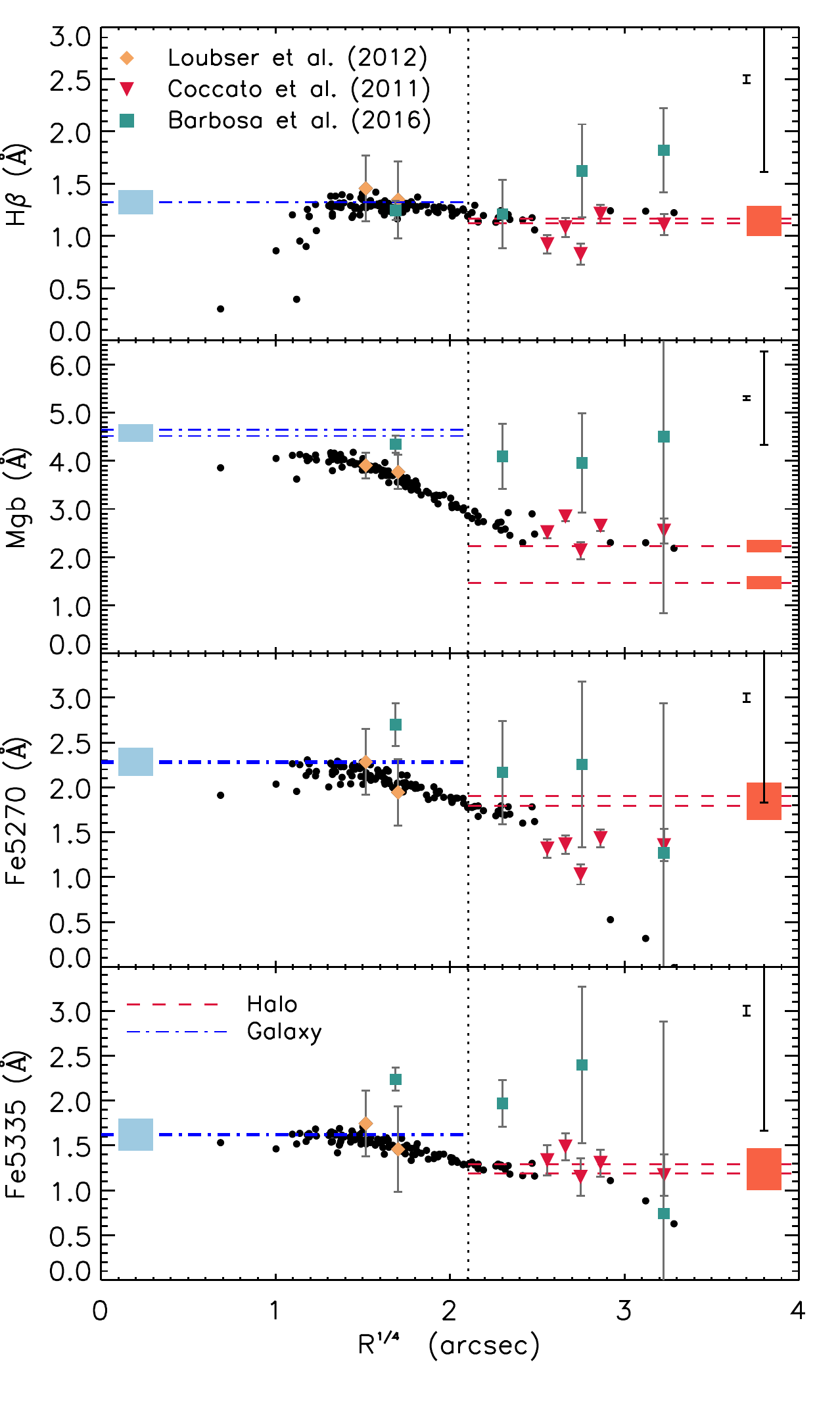}
\caption{
Absorption line strengths as a function of radius. From top to bottom are H$\beta$, Mgb, Fe$_{5270}$ and Fe$_{5335}$. The small black points represent measurements from the binned datacube after normalising the velocity dispersion, and the vertical dotted line shows the radius at which the light profile switches from galaxy-dominated to halo dominated in Model~2. The blue and red horizontal dashed lines represent the line-index measurements for the galaxy and halo spectra respectively for both models, plotted in the regions where that component dominates the light from the galaxy and with their uncertainties marked as solid colours at the end of each line. The mean uncertainty in the measurements for the binned spectra are shown in the top-right corner, with the left and right error bars representing the uncertainties in the measurements at R$^{1/4} < 2.7$ and R$^{1/4} \ge 2.7$ respectively. A comparison with the literature results is shown by the coloured points: the yellow diamonds represent the data published by \citet{Loubser_2012} for the central regions of the galaxy, the red triangles represent the results of \citet{Coccato_2011b} for the halo-dominated region, and the green squares show the results across the galaxy from \citet{Barbosa_2016} after binning into steps of R$^{1/4}=0.5$ and with the error bars giving the standard deviation in the measurements.
\label{fig:compareobs}}
\end{figure}

The analysis of the formation histories of the galaxy and halo was carried out through a study of the light-weighted stellar populations within each component, as measured from their decomposed spectra. Combinations of the hydrogen (H$\beta$), iron (Fe5270 and Fe5335) and magnesium (Mgb) absorption line indices were used as indicators of age, metallicity and $\alpha$-element abundance. The line strengths were measured using the Lick/IDS index definitions \citep{Worthey_1994}, and the uncertainties in the line index measurements were estimated from the propagation of random errors and the effect of uncertainties in the line-of-sight velocities \citep{Cardiel_1998}. The line strength measurements are given in Table~\ref{Table:line_indices}, and plotted against the measurements of the corresponding line strengths in the radially binned spectra in Fig.~\ref{fig:compareobs}. The H$\beta$ line strength remains similar in the galaxy and halo model spectra, suggesting either little age difference or too old stellar populations to easily detect an age difference between these components. The magnesium and iron lines all show a drop in their values in the halo spectrum relative to the galaxy spectrum, indicating the presence of a metallicity difference between the components. In order to compare these results with a more traditional analysis, the kinematics-normalised datacube was binned to achieve a S/N ratio of at least 50, which was reduced to 10 in the outermost regions to allow us to extend the plot to these radii. Measurements of the line strengths were obtained in the same way, and plotted onto Fig.~\ref{fig:compareobs} as a function of radius from the centre of the galaxy. The trends seen in the radial line strengths generally reflect those between the galaxy and halo decomposed spectra, with the main offsets being the H$\beta$ measurements in the centre of the data cube and the iron lines in the outskirts of the halo. The low H$\beta$ measurements can be explained by a combination of the H$\beta$ emission in this part of the galaxy, which fills in the absorption lines at the same wavelength and thus lowers the absorption line strengths, and the superposition of galaxy and halo light in this part of the datacube. The low iron measurements are harder to understand, but a likely explanation would be that these features are very weak and become lost in the high noise in these spectra, as reflected by the larger error bars for the outermost data points in the top right of each plot.

\begin{table}
\centering
\begin{tabular}{ccccc }
                        & \multicolumn{2}{c}{Model 1} &  \multicolumn{2}{c}{Model 2} \\
                        & Galaxy       & Halo       &  Galaxy       & Halo       \\ 
\hline

\multicolumn{1}{l|}{H$\beta$ (\AA)}    &  1.32$\pm$0.12            & 1.12$\pm$0.12   & 1.32$\pm$0.11                   & 1.16$\pm$0.12               \\
\multicolumn{1}{l|}{Mgb (\AA)}         & 4.64$\pm$0.12            & 1.47$\pm$0.14     & 4.52$\pm$0.12                            & 2.23$\pm$0.13         \\
\multicolumn{1}{l|}{Fe$_{5270}$ (\AA)}      & 2.29$\pm$0.15            & 1.79$\pm$0.15      &  2.27$\pm$0.16        & 1.90$\pm$0.15     \\
\multicolumn{1}{l|}{Fe$_{5335}$ (\AA)}     &  1.63$\pm$0.18          & 1.19$\pm$0.19   &   1.61$\pm$0.17              & 1.29$\pm$0.18            \\  
\\
\multicolumn{1}{l|}{Age (Gyrs)}    &  15$\pm$3            & 15.85$^*$   & 15$\pm$3                   &15.85$^*$               \\
\multicolumn{1}{l|}{[M/H]$_{\text{[MgFe]}'}$}         & 0.24$\pm$0.24            & $-0.75^*$     & 0.22$\pm$0.24                            & $-0.45^*$        \\
\multicolumn{1}{l|}{[$\alpha$/Fe]}      & 0.50$\pm$0.10            & $-0.10\pm$0.20      &  0.50$\pm$0.10        & 0.20$\pm$0.15     \\
\multicolumn{1}{l|}{[M/H]$_{15Gyrs}$}     &  0.22$\pm$0.08          & $-0.89\pm$0.19   &   0.17$\pm$0.08              & $-0.54\pm$0.18            \\ \hline

\end{tabular}
\caption{Line strength measurements for the H$\beta$, Mgb, Fe5270 and Fe5335 features from the decomposed galaxy and halo spectra for each model, and the corresponding stellar populations derived from these measurements. The halo measurements marked with a $^*$ should be used with caution since they lie too far from the SSP model for a reliable interpolation (see Fig.~\ref{fig:stellar_pops}).}
\label{Table:line_indices}
\end{table}

At this point one should note that in the majority of cases, the galaxy and halo spectrum measurements are in agreement between the two models. The only exception in Fig.~\ref{fig:compareobs} is the Mgb line strength in the halo models, which will have an impact on the age, metallicity and $\alpha$-enhancement estimates of that component. Since the SED created using Model~1 was found to show a poorer fit to that part of the original galaxy spectrum in Fig.~\ref{fig:spectra}, it is likely that the region around the magnesium triplet was particularly poorly fit with this model. One possible reason for this poor fit include the higher S\'ersic index for the halo in Model~1 leading to more light from the wings of the halo being incorporated into the one-dimensional integrated spectrum, increasing the noise. Alternatively, as shown in Fig.~\ref{Fig:light_prof}, the higher S\'ersic index of the galaxy in this model has attributed some of the Mgb line strength in the halo to the galaxy, thus reducing the Mgb line strength in the halo and increasing it in the galaxy. Since the line strengths from the galaxy and halo spectra obtained using Model~1 more closely reflect those from the galaxy and halo-dominated regions of the datacube, we consider the results using the double S\'ersic model to more accurately reflect the properties or NGC~3311.

A comparison with the literature was carried out, using the results from \citet{Loubser_2012} for the central regions of the galaxy, \citet{Coccato_2011b} for the outer halo, and \citet{Barbosa_2016} over the whole galaxy$+$halo system. We compare our line-strength measurements with theirs in Fig.~\ref{fig:compareobs} after correcting to the same effective spectral resolution. In order to present the results cleanly, the data from \citet{Barbosa_2016} have been selected to include only the measurements for slits that fell within the MUSE observations used here, and these measurements were then binned according to radius from the centre of the galaxy. The median values for each bin are shown in Fig.~\ref{fig:compareobs}, with the error bars reflecting the standard deviation in the individual measurements. It can be seen that our model galaxy and halo measurements (horizontal lines) and the radial measurements from the binned spectra (black points) are in general agreement with those of \citet{Loubser_2012} within the galaxy-dominated regions, with  \citet{Coccato_2011b} in the halo dominated region, and with \citet{Barbosa_2016} throughout the galaxy, where the majority of our measurements are within 1--2~error bars of the literature values.

\subsection{Age and metallicity}\label{sec:age_met}

\begin{figure*}
  \includegraphics[width=0.9\linewidth]{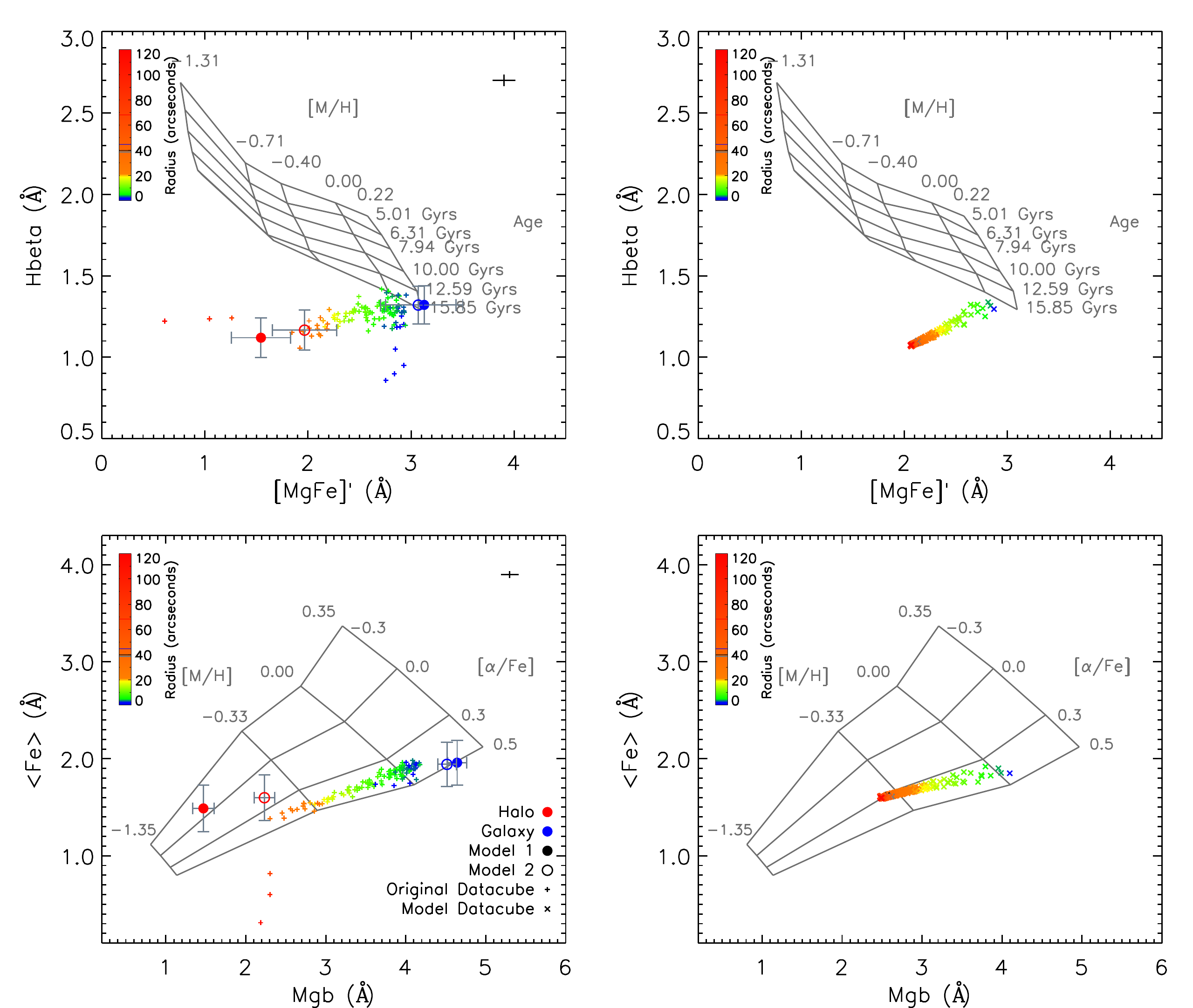}
  \caption{Lick index measurements from the decomposed galaxy and halo spectra superimposed upon SSP models for age and metallicity (top) and $\alpha$-element abundance (bottom). In the left column, large data points represent measurements from the decomposed spectra, with filled and hollow symbols representing Models 1 and 2 respectively. Small {\tiny+} points represent measurements from the original datacube, colour-coded as a function of radius as shown by the colour bars. In the right column, the small {\tiny $\times$} points represent measurements from the combined galaxy+halo model datacube for comparison, colour coded according to radius. The error bars in the top right corner of the left plots represent the mean error on the radial measurements. No error bars have been included in the plots on the right since the measurements on these plots are from the model datacubes, and therefore should have no random errors.
 \label{fig:stellar_pops}}
\end{figure*}

The age and metallicity of each component were determined by using the SSP models of \citet{Vazdekis_2010} to convert the line index measurements into estimates of the light-weighted ages and metallicities. The Single Stellar Population (SSP) models use the MILES stellar library \citep{Sanchez_2006}, which has a spectral resolution of $2.5\,$\AA\ (FWHM), and which were convolved with a Gaussian of the appropriate dispersion to reproduce the spectral resolution of the data. The resultant SSP models are thus matched to the data, minimizing the loss of information that normally occurs when degrading the data to match lower-resolution models. The H$\beta$ absorption feature was taken as a measure of age, and metallicity was measured using the combined metallicity index, 
\begin{equation} 
	\text{[MgFe]$'$}=\sqrt{\text{Mg}b\ (0.72 \times \text{Fe}5270 + 0.28 \times \text{Fe}5335)},
	\label{eq_MgFe}
\end{equation}
due to its negligible dependence on the $\alpha$-element abundance \citep{Gonzalez_1993,Thomas_2003}. The results from these measurements are given in Table~\ref{Table:line_indices} and in the top-left panel of Fig.~\ref{fig:stellar_pops} for the galaxy and halo model spectra (Models 1 and 2).

Interestingly, although the values of the indices corresponding to the galaxy models are compatible with those of old and relatively metal-rich stellar populations, the points corresponding to the halo models are clearly outside the model grid, with H$\beta$ values systematically below those of even the oldest stellar population values. We need to consider whether this systematic offset might somehow be an artefact of the fitting process. However, in this test case, the data are of sufficient quality to allow us to measure the indices directly from the raw on-sky spectra. As is apparent from Fig.~\ref{fig:stellar_pops}, this same offset is present in the radially-binned datacube in which the kinematics have been normalised (plotted as small points, colour coded to reflect the distance from the centre of the galaxy), which track consistently from galaxy to halo values with increasing radius, so the low values of H$\beta$ are clearly not a systematic error arising from the fitting process. 

We can also rule out that the low values of H$\beta$ are due to some systematic problem affecting the MUSE data because the completely independent datasets of \citet{Loubser_2012}, \citet{Coccato_2011b} and \citet{Barbosa_2016}, taken with different telescopes and instruments, and independently reduced, are compatible with our measurements and also fall outside the model grid.

We cannot but conclude that the problem must lie with the models. Very significant uncertainties still exist in producing reliable spectral models, in particular for stellar populations with non-solar metallicities and element ratios \citep[see, e.g.,][and references therein]{Thomas_2003,Thomas_2011,Vazdekis_2010}. Therefore, it is not surprising that the model predictions fail to reproduce the observations corresponding to stellar populations in the outer halos of galaxies, for which these models were never intended nor calibrated. 

Having said that, and despite all the possible shortcomings of the models, it seems reasonable to conclude that the stellar populations of the galaxy and the halo (and at all galactic radii) are very old, and probably older than $\sim$10\,Gyrs. Because virtually all age diagnostics saturate for such old ages, we cannot say whether an age gradient exists or not with any certainty. Neither can we rule out age differences of the order of a few billion years. 

The  [MgFe]$^\prime$ values shown in Fig.~\ref{fig:stellar_pops} suggest the existence of the kind of metallicity gradient with radius created by the observed light being dominated by the galaxy light at small radii\footnote{Note that the $H\beta$ values of the four or so innermost (bluest) points in the top-left panel of in Fig.~\ref{fig:stellar_pops} correspond to the central $\sim$4~arcsec of the galaxy, where line emission partially fills in the H$\beta$ absorption feature and skews the line-index measurements to lower values. During the fitting process this region was masked out, and so the analysis presented here is robust against contamination from the emission features.} and transitioning to halo-dominated light at larger radii.
The colour bar reflects this transition, such that the yellow  datapoints represent those measurements at $\sim$19.6~arcseconds, the radius at which the halo light starts to dominate the light profile for the double-S\'ersic model (Model 2) at $\sim$6000\AA. It is important to note at this point that the line strength measurements for the three outermost bins from the original datacube in both plots in Fig.~\ref{fig:stellar_pops} lie offset from the trends created by the radially binned spectra and the decomposed halo spectra. The same effect can be seen in Fig.~\ref{fig:compareobs} for the measurements of the iron indices. As described in Section~\ref{sec:stellar_pops}, these spectra were binned to achieve a lower S/N ratio of 10, and have correspondingly larger uncertainties which were not included in the error bars in  Fig.~\ref{fig:stellar_pops}. These measurements have therefore been included for completeness, but may not accurately represent the stellar populations in the halo-dominated region.

\textsc{buddi} also produces the model datacubes for each component, allowing the user to measure gradients across the whole galaxy and within the galaxy and halo components independently. No significant spectral gradients were detected within the galaxy and halo model datacubes, while Fig.~\ref{fig:stellar_pops} (top-right panel) shows that the combined galaxy+halo model datacube cleanly reproduces the trend seen in the original datacube outside of the central masked region. This trend suggests that the galaxy and halo contain distinct stellar populations, and thus that they may have been created in a two-step process. At this point it is interesting to note that the measurements for the decomposed galaxy spectra are slightly offset from the measurements at the centre of both the model and original datacubes. The most likely explanation for this offset is the superposition of halo light on top of the galaxy in that part of the datacube, leading to a systematic offset in the line strength measurements which is absent in the pure-galaxy spectrum. This effect represents the advantage gained by using \textsc{buddi} to cleanly extract spectra from multiple galaxy components and study their stellar populations with minimal contamination from other structures present.

Together, these measurements fit well with a scenario in which all the stars that make up this system formed very early, with the halo containing significantly more metal-poor stars than the main body of the galaxy. This could be naturally explained if the halo stars were formed in less massive galaxies that merged with a more massive and metal-rich progenitor. We will re-examine and quantify this hypothesis later.  

The trends between the galaxy and halo spectra presented in this section do not change significantly whether we use Model~1 (de Vaucouleurs + exponential) or Model 2 (double S\'ersic), indicating that the additional constraints in the fit parameters of Model~1 have little effect on the derived spectra, and that our conclusions do not depend sensitively on the model chosen. However, care should be taken when deriving the relative ages and metallicities using different models, as additional constraints may affect the resultant spectra, as show here by the variation in the Mgb line strength of the halo spectrum from the two models.

\subsection{Metallicity and $\alpha$-element abundance}\label{sec:alpha-Fe}

While the light-weighted age measurements primarily tell us about how long ago the last episode of star formation occurred, the $\alpha$-element abundance can provide information about the star-formation timescale. $\alpha$-element enhancement occurs when star formation is complete on a timescale shorter than the point at which Type Ia supernovae start to occur ($\sim$1\,Gyr). In that case, the main source of heavy elements recycled into stars are Type II supernovae that produce $\alpha$-elements in relatively high abundance with respect to Fe-peak elements. Thus, a higher $\alpha$-enhancement would reflect a shorter star-formation timescale.

The $\alpha$-enhancement of the central galaxy and halo are shown on the bottom-left of Fig.~\ref{fig:stellar_pops}. The iron index was calculated as%
\begin{equation} 
\langle\text{Fe}\rangle=(\text{Fe}5270 + \text{Fe}5335)/2,
\end{equation}
and was plotted against magnesium on the stellar population models of \citet{Thomas_2011}  for solar-normalised logarithmic $[\alpha/{\rm Fe}]$ values of $-0.3$, $0.0$, $0.3$ and $0.5$, with metallicities $[{\rm M}/{\rm H}]=\log(Z/Z_\odot)$ ranging between $-1.35$ and $+0.35$ for an age of $15\,$Gyrs\footnote{Our conclusions would be the same for any stellar population older than $\sim10\,$Gyrs.}. Since the models were measured at the Lick resolution, they were broadened using the same transformations used previously for the literature results. The radial measurements from the original kinematics-normalised datacube and from the model datacube are also plotted as before.

When comparing the results derived using the double S\'ersic model, the decomposed galaxy spectrum appears to be slightly more $\alpha$-enhanced ($[\alpha/{\rm Fe}]\sim0.5$) than the decomposed halo spectrum ($[\alpha/{\rm Fe}]\sim0.2$). According to \citet{Bender_2015}, this trend provides further evidence for the early, rapid formation of all the stars in the main body of the galaxy, while the star-formation timescale for the halo stars was probably longer. However, it must be noted that this trend is weak due to the size of the uncertainty on each measurement, and was not seen in earlier works by  \citet{Loubser_2012}, \citet{Coccato_2011b} and \citet{Barbosa_2016}, and so care must be taken when using this measurement alone to derive the origins of the galaxy and halo.

As discussed in Section~\ref{sec:age_met}, Fig.~\ref{fig:stellar_pops} also suggests that the decomposed galaxy spectrum corresponds to a significantly more metal-rich stellar population than that of the halo spectrum. Given the mismatch between models and observations, any absolute metallicity value should be taken with great caution. Nevertheless, it is not unreasonable to expect relative values to be, perhaps, more reliable. If that is the case, the main body of the galaxy would be $\sim$0.6~dex more metal rich than the halo. If we interpret this difference according to the hypothesis that cD galaxies originated as normal ellipticals whose extended stellar halos have grown due to mergers \citep[e.g.,][]{Bender_2015,Zhao_2015b}, the mass--metallicity relation would imply that the merging galaxies that built the halo were much less massive than the original central elliptical 

The $z\simeq0$ mass--metallicity relation of \citet{Ma_2016}, 
\begin{equation} 
\log(Z/Z_\odot) = 0.4\log (M/M_\odot)-4.37,
\end{equation}
can be used to estimate that the galaxies that built the halo of this cD galaxy were, on average, $\sim50$ times less massive than the progenitor elliptical galaxy. In other words, under this hypothesis, the cD halo grew through minor mergers. This seems at odds with some theoretical expectations \citep[][and references therein]{Nipoti_2017} which suggest that the merging galaxies should be relatively massive since dynamical friction's efficiency increases linearly with galaxy mass. However, the merging histories of BCGs are expected to be very varied \citep{DeLucia_2007}.

Our results also suggest that both the central elliptical and the merging galaxies formed the bulk of theirs stars more than $\sim10\,$Gyrs ago (at $z\gtrsim2$), but, while the majority of the stars in the central galaxy formed over a very short timescale ($\ll1\,$Gyr), the star formation in the merging galaxies was much more extended ($\gtrsim1\,$Gyr).

\section{Conclusion}\label{sec:conclusion}

In this paper, we have presented a pilot study of integral field spectroscopic observations of the cD galaxy NGC~3311 to demonstrate that such systems can be decomposed into the main galaxy component and the surrounding stellar halo.  The fits were carried out with \textsc{buddi}, using both a de Vaucouleurs$+$exponential  and a double S\'ersic profile to model the galaxy and its halo. Despite the limited spatial coverage of the galaxy, both models were found to fit the galaxy and halo well, with the main difference being the treatment of the light profile in the central masked region and outer halo outside of the field of view of the data. While the best fit to the galaxy SED was achieved using the double S\'ersic model, the conclusions of our study are very similar for both models, giving some confidence that the results do not depend over-sensitively on such details. The stellar populations analysis on the decomposed spectra for each component found that the bulk of the stars in both the central galaxy and its extended halo are very old, with the galaxy being significantly more metal rich and more $\alpha$-enhanced than the halo.  It is noteworthy that the apparent gradients in age and metallicity indicators with distance from the centre of the galaxy are entirely consistent with the radially-varying contributions of galaxy and halo components, which individually display no gradients. The high $[\alpha/{\rm Fe}]$ ratio and metallicity of the central galaxy suggest that its stars were formed very rapidly in a massive galaxy through a quick, gas-rich dissipative collapse at high redshift ($z\gtrsim2$). The lower metallicity of the halo indicates that it has been built up through the accretion of much less massive galaxies. These smaller galaxies that later contributed to the halo were also built relatively early on, perhaps originating as disk galaxies that formed stars in a more steady manner until the star formation was quenched by the cluster environment on a longer timescale. This formation scenario is consistent with those presented previously for  NGC~3311 by \citet{Coccato_2011b}, \citet{Arnaboldi_2012} and \citet{Barbosa_2016, Barbosa_2018}, and those seen in other cD galaxies \citep[e.g. ][]{Bender_2015, Zhao_2015b, Longobardi_2015}.

The consistency between the results from the decomposed spectra with those measured through a more traditional radial analysis of the datacube indicate that this technique could be applied to smaller, lower-quality data cubes where we do not have enough signal to measure the gradients in spectral properties explicitly. Additionally, the technique could be applied to more extensive survey data in the coming years to determine how the formation histories of the extended stellar halos  around cD galaxies might depend on other properties of these systems.

\section*{Acknowledgements}

We would  like to thank the anonymous referee for their useful comments that helped improve this paper. Based on observations collected at the European Organisation for Astronomical Research in the Southern Hemisphere under ESO programme P094.B-0711 (PI: Arnaboldi). This research has made use of the NASA/IPAC Extragalactic Database (NED) which is operated by the Jet Propulsion Laboratory, California Institute of Technology, under contract with the National Aeronautics and Space Administration.

%

\end{document}